# Beyond a phenomenological description of magnetostriction

A. H. Reid[1,2]*, X. Shen[3], P. Maldonado[4], T. Chase[1,5], E. Jal[1], P. W. Granitzka[1,6], K. Carva[7], R. K. Li[3], J. Li[8], L. Wu[8], T. Vecchione[3], T. Liu[1,9], Z. Chen[1,9], D.J. Higley[1,5], N. Hartmann[2], R. Coffee[2], J. Wu[3], G.L. Dakowski[2], W. Schlotter[2], H. Ohldag[10], Y.K. Takahashi[11], V. Mehta[12,13], O. Hellwig[12,14,15], A. Fry[2], Y. Zhu[8], J. Cao[16], E.E. Fullerton[17], J. Stöhr[1], P. M. Oppeneer[4], X.J. Wang[3] & H.A. Dürr[1]*

[1] Stanford Institute for Materials and Energy Sciences, SLAC National Accelerator Laboratory, 2575 Sand Hill Road, Menlo Park, CA 94025, USA.

[2] Linac Coherent Light Source, SLAC National Accelerator Laboratory, 2575 Sand Hill Road, Menlo Park, CA 94025, USA.

[3] Accelerator Division, SLAC National Accelerator Laboratory, 2575 Sand Hill Road, Menlo Park, CA 94025, USA.

[4] Department of Physics and Astronomy, Uppsala University, P. O. Box 516, S-75120 Uppsala, Sweden

[5] Department of Applied Physics, Stanford University, Stanford, California 94305, USA

[6] Van der Waals-Zeeman Institute, University of Amsterdam, 1018XE Amsterdam, The Netherlands

[7] Charles University, Faculty of Mathematics and Physics, Department of Condensed Matter Physics, Ke Karlovu 5, CZ-12116 Prague 2, Czech Republic

[8] Brookhaven National Laboratory, Upton, New York 1193, USA

[9] Department of Physics, Stanford University, Stanford, California 94305, USA

[10] Stanford Synchrotron Radiation Laboratory, SLAC National Accelerator Laboratory, 2575 Sand Hill Road, Menlo Park, CA 94025, USA.

[11] Magnetic Materials Unit, National Institute for Materials Science, Tsukuba 305-0047, Japan

[12] San Jose Research Center, HGST a Western Digital company, 3403 Yerba Buena Road, San Jose, California 95135, USA.

[13] Thomas J. Watson Research Center, 1101 Kitchawan Road, Yorktown Heights, New York 10598, USA.

[14] Institute of Physics, Technische Universität Chemnitz, Reichenhainer Straße 70, D-09107 Chemnitz, Germany

[15] Institute of Ion Beam Physics and Materials Research, Helmholtz-Zentrum Dresden–Rossendorf, 01328 Dresden, Germany.

[16] Department of Physics and National High Magnetic Field Laboratory, Florida State University, Tallahassee, Florida 32310, USA

[17] Center for Memory and Recording Research, UC San Diego, 9500 Gilman Drive, La Jolla, CA, 92093-0401, USA

*email: alexhmr@slac.stanford.edu; hdurr@slac.stanford.edu

**Abstract**

**Magnetostriction, the strain induced by a change in magnetization, is a universal effect in magnetic materials. Owing to the difficulty in unravelling its microscopic origin, it has been largely treated phenomenologically. We show how the source of magnetostriction—the underlying magnetoelastic stress—can be separated in the time domain, opening the door for an atomistic understanding. X-ray and electron diffraction is used to separate the sub-picosecond spin and lattice responses of FePt nanoparticles. Following excitation with a 50-fs laser pulse, time-resolved X-ray diffraction demonstrates that magnetic order is lost within the nanoparticles with a time constant of 146 fs. Ultrafast electron diffraction reveals that this demagnetization is followed by an anisotropic, three-dimensional lattice motion. Analysis of the size, speed and symmetry of the lattice motion, together with *ab initio* calculation accounting for the stresses due to electrons and phonons, allow us to reveal the magnetoelastic stress generated by demagnetization.**

The functional properties of materials often depend on the detailed and subtle interplay of electronic, spin and lattice degrees of freedom.[1–16] The complexity of this interplay can leads to a variety of technologically useful behaviors. These effects include anomalous thermal expansion,[1,10,11] optical switching of magnetization,[11–17] and superconductivity.[18] Understanding the details of this electron–spin–lattice interplay remains one of the most challenging scientific problems in condensed matter physics. A particularly perplexing aspect is the strong coupling of magnetic spin to electron and lattice degrees of freedom observed in magnetically ordered metals. In such metals, spin order can be extinguished on a timescale of order 100 fs.[4-8] This fast timescale implies strong coupling of spin to the electronic system, seemingly not limited by the requirement of angular momentum conservation that necessitates the involvement of the lattice. This observation poses an interesting question, how strong or fast are the interactions that govern the spin-lattice effect of magnetostriction.

Iron platinum (FePt) alloys are a particularly rich example of materials that exhibit such electron–spin–lattice interplay. In the $L1_0$ crystal phase, FePt is ferromagnetically ordered with extremely high magneto-crystalline anisotropy, making FePt the material of choice for next-generation magnetic storage media.[19,20] In addition, the FePt system displays magnetostriction, leading to anomalous thermal expansion with a *c*-axis contraction in the paramagnetic phase.[21] The anomalous thermal expansion leads to a temperature-dependent change in the tetragonal lattice distortion, a quantity that is instrumental for establishing the high magneto-crystalline anisotropy in FePt.[22] Recent reports have shown that FePt also exhibits an all-optical magnetization reversal when subjected to ultrashort pulses of circularly polarized optical light.[13,14] However, the mechanism of how light switches magnetization in FePt is not yet understood. Therefore, understanding the interplay of electrons, spins, and lattice in FePt is a necessary step towards unraveling the mysteries of its rich functional properties.

Experimentally, lattice stress is not directly observable. Instead, measurements can be made of lattice strain – the physical movement of atoms in response to the stress. The strain response is often complex in solids where macroscopic effects can constrain the motion. For this reason, we choose nanoparticle grains of FePt held in a free-standing carbon matrix (see Fig.1a) as the medium to understand the stress. Previous studies of magnetic structural dynamics were performed on continuous thin films deposited on substrates.[23-25] Our use of unconstrained FePt single-crystal nanoparticles allows us to study the full 3-dimensional lattice motion on the natural timescale of the strain propagation through the nanoparticles.[26,27] This approach separates the individual contributions from electrons, spins, and phonons to lattice stress in FePt particles via their different symmetry properties and temporal onsets, as illustrated in Fig.1b. By employing *ab initio* calculations to capture the non-equilibrium stresses of electrons and phonons, we show that a large magnetoelastic stress term related to magnetostrictive spin–lattice coupling must contribute to the observed strain.[3] This term dominates the anomalous lattice expansion of FePt under the non-equilibrium condition during the first few ps following fs laser excitation.

**Results**

In Fig. 1a we show a TEM image of the freestanding sheet of FePt nanoparticles embedded in carbon which is used in this study. The FePt nanoparticles are approximately cylindrically shaped with large surfaces that are free to respond to dynamic stresses along the surface normal. The particles are grown to have their magnetic easy axis, the crystallographic *c*-axis, along the surface normal (see Methods). We expect that a linear stress-strain surface response will dominate the motion for a time of $\tau = r_{np}/v_s$, where $r_{np}$ is the nanoparticle radius (mean $r_{np} = 6.5$ nm) and $v_s$ is the speed of sound. A value of $\tau \approx 3$ ps is estimated for these FePt nanoparticles using the FePt sound velocity of 2.2 nm/ps.[22]

We measured the spin and lattice responses separately on the same FePt-C samples (see Methods), although using different experimental setups. Figure 2 shows the spin and lattice responses of a free-standing layer of FePt nanoparticles to fs laser excitation. We measured the spin response with ultrafast X-ray resonant diffraction at the Fe $L_3$-edge using circularly polarized X-rays.[15] Figure 2b shows the measured Fe spin response following fs laser excitation (see schematic in Fig. 2a & Methods). The data shown in Fig. 2b is the difference of two measurements of the resonant X-ray diffraction intensity with the sample magnetized in opposite directions by applying an external magnetic field of ±0.4 T during the experiment. This field strength is significantly lower than the field required to switch the FePt nanoparticles when no laser pump is applied; the switching field without laser pump at room temperature is measured to be ±4 T.[28] The large reduction of the switching field indicates that the magneto-crystalline anisotropy barrier between opposite magnetization directions is strongly reduced by laser excitation. The data in Fig. 2b are fitted with a double exponential,[6] resulting in a time constant for demagnetization of 146 ± 15 fs, and a time constant of 16 ± 4 ps for the subsequent magnetization recovery. These results agree with optical fs laser experiments reported earlier.[6,7] The fitted demagnetization amplitudes vs. laser pump fluence are shown in the inset of Fig. 2b.

We use ultrafast electron diffraction (UED) to directly measure the FePt nanoparticles' lattice response.[29,30] The nanoparticles were crystallographically aligned from their growth on a single-crystal substrate; the substrate was subsequently removed (see Methods). Therefore, the FePt diffraction pattern, shown in Fig. 2c, shows well-defined Bragg peaks. From time-resolved measurements of multiple Bragg diffraction peaks, the temporal evolutions of FePt lattice spacings along the *a* and *c* directions (the *b* direction evolves identically to the *a* direction) are extracted and shown as a function of time in Fig. 2d

(see Methods for details). We used a pump fluence of 5 mJ/cm$^2$ that corresponding to a fs demagnetization of 30% of the initial magnetization (see inset of Fig. 2b). The measured change in lattice spacing along the *c* direction is clearly different from the change along the *a*- and *b*-lattice directions. The lattice spacing in the *a* and *b* directions of FePt shows a rapid expansion which peaks at approximately 3 ps. The expansion is followed by an oscillatory motion. The dynamics along the *c* direction is very different. The *c*-lattice spacing shows a rapid initial contraction on a similar timescale to the *a*- and *b*-lattice expansions. This *c*-lattice motion also oscillates anti-phase to the *a* and *b* directions. It then relaxes back towards the pre-excitation value on a timescale of ~20 ps, whereas this does not occur in the *a* and *b* directions which show expansion.

**Discussion**

We are interested in understanding the dynamic response of the FePt lattice to the initial stresses generated during fs heating and demagnetization. To better visualize this initial lattice response, we replotted the FePt nanoparticle's unit-cell volume and its *c–a* lattice vectors for two laser fluences in Fig. 3. The non-equilibrium state of the system is modelled using four thermal baths for the occurring degrees of freedom of the nanoparticle. To this end, we extend the three-temperature model of FePt,[31] to include a temperature representing the carbon matrix (see Methods). The 4-temperature model results are shown in Fig. 3a. The temporal evolution of the unit-cell volume, V(t) = c(t) · a(t)$^2$ (note that a(t) = b(t)), is shown in Fig. 3b. This volume evolution shows that there is a clear cross-over from rapidly increasing volume to a steady state regime at around 3 ps; this cross-over coincides with the first extremum of the oscillation in Fig. 2d. The 3 ps cross-over can also be identified as the turning point of the *a–c* trajectory plot in Fig. 3b, which displays the relative lattice change in the *a–c* plane. The characteristic motions are clearly separated. Initially, the lattice system moves along a trajectory defined by the stresses at the nanoparticle surfaces towards a turning point at 3 ps time delay. This delay is defined by the time taken for the strain waves to reach the center of the nanoparticles. Following this time delay, the nanoparticles begin a damped ringing motion that brings them back to the quasi-equilibrium state for *c–a* thermal expansion (shown in blue as a function of laser fluence in Fig. 3b).

To understand the initial expansion trajectory in Fig. 3b we conducted *ab initio* calculations of the electron and phonon stress contributions.[32,33] The electronic stress $\sigma^e$ is evaluated from a calculation of the electronic Grüneisen parameter along different symmetry directions (see Supplementary Materials for details). We determined that the electronic stress is anisotropic with $\sigma_{a,b}^e = 2\sigma_c^e$, but has a positive electronic pressure along all crystallographic axes. Next, to treat the phonon stress contribution, we calculated the out-of-equilibrium behavior of the lattice from the phonon populations, assuming independent phonon modes and including phonon–phonon interactions (see Methods). By UED measurement of the Debye–Waller effect, we experimentally determine that these modes are populated exponentially with a time constant of 2.7 ps.[34] The *ab initio* calculated mode-dependent contributions to the phonon stress are shown in Fig. 4a. As with the electronic stress, we find that the non-equilibrium phonon stress is highly anisotropic with $\sigma_{a,b}^{ph} = 7.3\sigma_c^{ph}$, but it remains positive for all crystallographic directions. Consequently, to explain the strong negative strain observed along the *c* axis, a further stress contribution arising from the magnetic system must be considered.

Any *ab initio* calculation of the magnetic stress relies on the details of the coupling between spins and lattice, and, spins and electrons. For ultrafast demagnetization these details remain unknown. Instead we attempt a more robust approach. We determine the structural ground state of FePt in the ferromagnetic (FM) and paramagnetic (PM) phases using spin-polarized density functional theory in the local density approximation (see Methods for details). The calculated total energies for constant volume are shown as a function of the *c/a* ratio in Fig. 4b and are in good agreement with the experimental values. The difference between the two structural energy minima for FM and PM phases are characterized by a -0.53 % *c*-lattice contraction and +0.25 % *a*-lattice expansion. It is assumed that the straight line trajectory between these two phases characterizes the magnetic stress vector, i.e. $\sigma_{a,b}^m \approx -0.47\sigma_c^m$. This stress trajectory is represented by the green arrow in Fig. 3c, and shows the required negative stress along the *c* axis. This would be the trajectory predicted for a 100% demagnetization, under static equilibrium conditions. It is noted that in our experiment we have neither 100% demagnetization nor equilibrium conditions.

To understand the relative sizes of the different stress contributions in the experiment, we can consider the FePt lattice response in terms of a simple coupled harmonic oscillator with a 2-dimensional displacement coordinate: $Q_i$ ($i = a, c$).[35] The lattice motion in Fig. 2d is modeled by:

$$\frac{d^2 Q_i}{dt^2} + 2\beta \frac{dQ_i}{dt} + \sum_j \Lambda_{ij} Q_j = \sigma_i(t) \qquad (1)$$

where $\beta$ is a phenomenological damping constant and, $\Lambda_{ij}$ is a matrix whose diagonal terms describe the frequency, $\omega$, of the coherent breathing mode, while its off-diagonal terms represent the elastic coupling between $j = ab, c$ lattice strains. Equation (1) is characterized by two solutions, with symmetric ($Q_{a,b} \sim Q_c$) and antisymmetric ($Q_{a,b} \sim - Q_c$) eigenvectors and eigenfrequencies $\omega_1 = \pm\sqrt{\Lambda_{aa} + 2\Lambda_{ac}}$ and $\omega_2 = \pm\sqrt{\Lambda_{aa}}$, respectively (we assume here that $\Lambda_{aa} = \Lambda_{cc}$). Fig. 2d shows that the antisymmetric solution is favored. We note that the symmetry of the resulting lattice strain amplitude will depend on the symmetry of the driving stress terms as previously discussed. We describe the driving stresses in equation (1) as:

$$\sigma_{a,b}(t) = \sigma_{a,b}^e + \sigma_{a,b}^m + \sigma_{a,b}^{ph} + \sigma_{a,b}^{carb},$$

$$\sigma_c(t) = \sigma_c^e + \sigma_c^m + \sigma_c^{ph} \quad (2),$$

In the following paragraphs, we detail the form of the stresses assumed in the harmonic oscillator model.

As discussed in the supplementary information the electronic stress $\sigma_i^e$ is determined by the electronic temperature and electronic heat capacity of the system: $\sigma_i^e = -\gamma_i^e C_e(T_e)\delta T_e$. The transient electronic temperature is determined by the 4-temperature model, while our *ab initio* electronic structure calculations determine the ratio of the coefficients to be: $\gamma_{ab} = 2\gamma_c$. The stress thus has the form: $\sigma_i^e = \sigma^e(\overleftarrow{ab} + 0.5\overleftarrow{c})T_e(t)\delta T_e(t)$. A single scaling constant, $\sigma^e$, is used to characterize the transient electronic stress.

The magnetic stress, $\sigma_i^m$, is predicted by our calculation to have the symmetry $2.12\sigma_{ab}^m = -\sigma_c^m$. We assume this to be the case, and, further, assume that the stress must be proportional to the square of the change in magnetization, due to time reversal symmetry. Therefore, the evolution of this stress, is determined by the measured change in the FePt magnetization, having the form: $\sigma_i^m = \sigma^m(0.47\overleftarrow{ab} - \overleftarrow{c})(\Delta M(t)/M_0)^2$, where $M(t)$ is defined by the bi-exponential fit to Fig 2a. Again, a single parameter, $\sigma^m$, is used to characterize the strength of this stress component.

The stresses due to phonons are dominated by the low-energy modes due to the larger atomic displacement per unit energy associated with these modes. For this reason, the phonon stress is not adequately modelled by the lattice temperature – a measure of the energy in the lattice – but must also account for the phonon thermalization time. The attenuation of the Bragg reflections due to the Debye—Waller effect is measured to determine a lattice equilibration time. It is found that the mean-square atomic displacements increase with an exponential time constant $\tau_l$ of 2.7 ps. Again, we use our *ab initio* theory calculations to determine the relative strength of the phonon stresses in the *a,b* & *c* directions ($\sigma_{ab}^{ph} = 7.3\sigma_c^{ph}$). The phonon stress thus has the form: $\sigma_i^{ph} = \sigma^{ph}(\overleftarrow{ab} + 0.14\overleftarrow{c})[1 - \exp(-\frac{t}{\tau_l})]$.

Finally, we note that the carbon matrix must resist the outward motion of the crystal in the *a,b* direction. The stress is assumed to evolve with the temperature of the carbon bath: $\sigma_{ab}^{carb} = \sigma^{carb}T_{carb}(t)$. No stress from the carbon is considered in the *c* direction, as nothing restrains the motion in this direction.

Using the above stress models, a least-squares non-linear curve fitting is made to the experimental data. The best fit to the experiment is shown in Fig. 2d. The form of these stresses present and the fitting values obtained are presented in Table 1. The results indicate that a non-zero magnetoelastic stress does develop within the FePt nanoparticle on the timescale of the ultrafast demagnetization. It is primarily this stress that drives in the anisotropic lattice displacement which proceeds as a strain wave moving inwards from the nanoparticple's boundary.

Our results demonstrate how the stress contributions for different degrees of freedom can be seperated in the time domain by measuring nanometer sized particles. In particular, we uncovered the existence of the magnetostrictive driving force for strongly anisotropic lattice motion in FePt nanoparticles. Magnetoelastic stress builds up on the sub-ps timescale, characteristic of ultrafast demagnetization. On the ps timescale, stress from transiently populated phonons takes over and results in a lattice tetragonality. Studies by Lukashev *et al.* have shown that a reduction in tetragonality favors a reduced FePt magnetocrystalline anisotropy barrier.[36] We speculate that this reduced barrier results in reduced magnetic coercivity, which allows laser-excited FePt nanoparticles to reverse their magnetization in a magnetic field of 0.4 T. This observation provides a new avenue towards manipulating the magnetocrystalline anisotropy in future laser-assisted magnetic data storage technologies and opens the possibility of using this ultrafast magnetostriction for new types of THz frequency magnetostrictive actuators.

**Methods**

Sample Preparation and Characterization

The single crystalline $L1_0$ FePt grains were grown epitaxially onto a single-crystal MgO(001) substrate by co-sputtering Fe, Pt and C.[28] This resulted in FePt nanoparticles of approximately cylindrical shape with heights of 10 nm and diameters in the range of 8 to 24 nm, with an average of 13 nm. The FePt nanoparticles form with *a* and *b* crystallographic directions oriented parallel to the MgO surface. The space in-between the nanoparticles is filled with glassy carbon, which makes up 30% of the film's volume. Following the sputtering process, the MgO substrate was chemically removed and the FePt-C films were floated onto copper wire mesh grids with 100-µm-wide openings. Figure 1a shows that individual nanoparticles remained aligned during this process, as the particles were held in place by the carbon matrix. The nanoparticles were characterized using transmission electron microscopy (see Fig. 1a). Temperature-dependent studies, shown in supplementary Fig. 3, were used to deduce an increase in *a/c* by 0.6 % together with a +0.3% *a*-axis expansion between 300 and 500 K.

Resonant magnetic X-ray scattering from granular FePt

The dynamic magnetic response of the FePt nanoparticles to laser pulses of 50 fs duration and 800 nm central wavelength was measured at the SXR instrument of the Linac Coherent Light Source (LCLS) at SLAC. The average optical absorption of these nanoparticles was previous calculated to be 21%.[16] We probed the magnetization changes using circularly polarized X-rays, with the X-ray energy tuned to the Fe 2p-3d resonance (708 eV photon energy).[15] Scattered X-rays were measured in transmission geometry by a pnCCD X-ray camera (illustrated in Fig. 2a). Radial scattering profiles after azimuthal angular averaging of the pnCCD patterns display a peak due to the interparticle scattering. Figure 2b displays the difference in X-ray scattering yield with the sample magnetization aligned in externally applied magnetic fields of ±0.4 T. This difference signal is proportional to the average particle magnetization along the X-ray incidence direction, switched by the external field. Note that 0.4 T is significantly lower than the static coercive field of 4 T in FePt nanoparticles, but allows a nearly 90% reversal of magnetization. However, we found that fs laser excitation, with the fluences shown in Fig. 2b, enabled a magnetization reversal similar to heat-assisted magnetic recording. We also measured the ultrafast magnetization dynamics of MgO supported FePt nanoparticles and found identical results to the freestanding FePt granular films shown in Fig. 2b.

Separating *c*-axis and *a*- *b*-axis motion in UED data

The dynamic response of the FePt lattice was measured by ultrafast electron diffraction in a transmission geometry with 3 MeV electrons from the SLAC ultrafast electron diffraction source.[30] The FePt nanoparticles were dynamically heated with a 1.55 eV, 50 fs optical laser pulse. To meet the Bragg condition for different lattice reflections, the film was rotated around axes normal to the probe beam. Due to geometrical restrictions, rotation angles were limited to 45° from normal incidence. Measurements were made at different incidence angles to access the Bragg peaks displayed in Supplementary Fig 2 in the Supplementary Information. The time evolution data for Bragg peak positions (*hkl*) shows large differences for peaks with different "*l*" indexes, i.e. different projections along the out-of-plane (*c*-axis) direction. Peaks with the same relative projection along the c axis, $l^2/(h^2 + k^2 + l^2)$, showed the same temporal response. Measurements of multiple Bragg reflections with different *c*-axis projections were used to extrapolate the motion along the *c*-axis (see Supplementary Fig. 2). The extrapolated motion for the FePt *c*-axis is shown in Fig. 2d, together with the directly measured *a*-axis (& *b*) motion.

*Ab-initio* calculations of electronic and phononic stress

The ultrafast laser excitation of the FePt nanoparticles is followed by a dynamical response of the material lattice which can be characterized by magnetic, electronic and phononic stresses. To achieve a complete determination of the latter two, we used classical kinetic theory along with the Fermi's golden rule of scattering theory to derive novel out-of-equilibrium rate equations for the electronic and phononic distributions.[37] We solve those equations by using input from first-principles calculations; where the incorporation of a temperature- and mode-dependent electron-phonon coupling and anharmonicities through phonon-phonon interaction are essential to attain a full solution of the time evolution. Subsequently, we obtained the induced stresses directly from the *ab initio* phononic and electronic distributions via a proportionality relation (see Supplementary Information on *ab initio* calculations). We determined the contribution along the different real-space directions by the mode and branch dependence of the Grüneisein parameter and the phononic distribution.

*Ab-initio* calculations of FePt ground state properties

The ground state properties of ferromagnetic FePt have been calculated using spin polarized density functional theory (DFT) in the local-density approximation (LDA) as implemented in the code ABINIT.[38]

The electron-ionic core interaction on the valence electrons was represented by projector-augmented wave potentials (PAW),[39] and the wave functions were expanded in plane waves with an energy cutoff at 29 Hartree and a cutoff for the double grid of 31 Hartree, which was sufficient to converge the total energy for a dense *k*-point sampling. Reciprocal space sampling was performed using the Monkhorst-Pack scheme with a *k*-point mesh of 32x32x32. After optimization, the lattice parameters for the FePt $L1_0$ (P4/mmm) structure were found to be $a=3.857$ Å and $c=3.761$ Å, comparable to the experimental values $a=3.852$ Å and $c=3.713$ Å. The resulting value of $c/a = 0.975$ is in good agreement with experiment ($c/a=0.964$, see Fig. S3c). The local magnetic moments of the Fe atoms in the ordered Fe-Pt structures in the ferromagnetic phase are 3.065 $\mu_B$.

For the paramagnetic (PM) state, we adopt the disordered local moments (DLM) approach, which states that paramagnetism can be modeled as a state where atomic magnetic moments are randomly oriented (i.e., non-collinear magnetism). The DLM approach can be simplified by considering only collinear magnetic moments when the spin-orbit coupling is not taken into account. Hence, the problem of modeling paramagnetism becomes a problem of modeling random distributions of collinear spin components. It can be solved by using special quasi-random structures (SQS). A SQS is a specially designed supercell built of ideal lattice sites to mimic the most relevant pair and multisite correlation functions of a completely disordered phase (PM order in our case). As PM simulation cell we adopted an extended lattice cell of 32 atoms (16 FePt unit cells). The disordered local moment approach (or SQS) provides a better description of the PM phase than the spin-non-polarized calculation, even without SOC, and gives good results for magnetostriction (even without SO).[40]

The thus-calculated PM phase shows an expansion of 0.25 % in the *a,b*-direction and a -0.53 % reduction of the *c*-lattice constant when moving from the ferromagnetic to the paramagnetic phase. Thus, the computed *c/a*-ratio of paramagnetic FePt becomes 0.964. The expected change in the lattice constants between the FM and PM phases under static conditions is indicated by the dashed green arrow in Fig. 3c.

Four-temperature model of non-equilibrium state

The microscopic three-temperature model (M3TM) has been developed to understand the nature and evolution of the non-equilibrium state in laser-excited ferromagnets.[41] This model has be recently applied to FePt:Cu by Kimling *et al.*[31] Here we adapt the model of Kimling *et al.* to our sample of FePt nanoparticles in a carbon matrix. We model the system as four coupled thermal baths: the FePt electronic bath; the FePt spin bath; the FePt phonon bath, and the carbon matrix bath. Each bath has its own associated temperature. Table 2 summarizes the parameters used in the model, the heat capacities of the four baths and relative coupling strengths between the baths. The four equations of the model are:

$$C_e \frac{dT_e}{dt} = g_{eph}(T_{ph} - T_e) + g_{es}(T_s - T_e),$$

$$C_{ph} \frac{dT_{ph}}{dt} = g_{eph}(T_e - T_{ph}) + g_{phc}(T_{carb} - T_{ph}),$$

$$C_s \frac{dT_s}{dt} = g_{es}(T_e - T_s),$$

$$C_{carb} \frac{dT_{carb}}{dt} = g_{phc}(T_{ph} - T_{carb}) \qquad (2),$$

The model is solved numerically for an excitation of the electronic temperature with a duration of 50 fs. The solution obtained is shown in Fig 3a. In the below section we explain how the 4-temperature-model of the non-equilibrium is used to model the non-equilibrium stresses in the harmonic oscillator model.

**Data availability**

The authors declare that the data supporting the findings of this study are available within the article and its Supplementary Information files. All other relevant data supporting the findings of this study are available on request.

**Acknowledgments:** The technical support from SLAC Accelerator Directorate, Technology Innovation Directorate, LCLS laser division and Test Facility Division is gratefully acknowledged. We thank R.K. Jobe, D. McCormick, A. Mitra, S. Carron and J. Corbett for their invaluable help and technical assistance. Research at SLAC was supported through the SIMES Institute which like the LCLS and SSRL user facilities is funded by the Office of Basic Energy Sciences of the U.S. Department of Energy under Contract No. DE-AC02- 76SF00515. Use of the ultrafast Electron diffraction facility at SLAC is supported by DOE BES, Scientific User Facilities Division, Accelerator and Detector R&D program and Laboratory Directed Research and Development funding under contract DE-AC02-76SF00515. P.M, K.C., and P.M.O. acknowledge support from the Swedish Research Council (VR), the Knut and Alice Wallenberg Foundation (grant No. 2015.0060), the Röntgen-Ångström Cluster, and the Swedish National Infrastructure for Computing (SNIC). Work at BNL was supported by DOE BES Materials Science and Engineering Division under Contract No: DE-AC02-98CH10886. J.C. would like to acknowledge the support from National Science Foundation Grant No. 1207252. E.E.F. would like to acknowledge support from the U.S. Department of Energy (DOE), Office of Basic Energy Sciences (BES) under Award No. DE-SC0003678.

**Author contributions:** A.H.R., H.A.D conceived the experiment, Y.K.T., E.E.F., O.H. grew and characterized the samples, A.H.R., X.S. T.C., R.K.L., T.V., N.H., R.C., J.W., X.W., H.A.D. performed the electron diffraction experiments and analysis, A.H.R., E.J., T.C., P.G., T.L., Z.C., D.H., G.D., W.S., H.O., J.S., H.A.D. performed the x-ray experiments and analysis, A.H.R., J.L., Y.Z., J.C. performed the simulations, P.M., K.C., P.O. performed *ab initio* calculations, A.H.R., H.A.D. coordinated writing of the paper with contributions from all coauthors.

**Figures and captions**

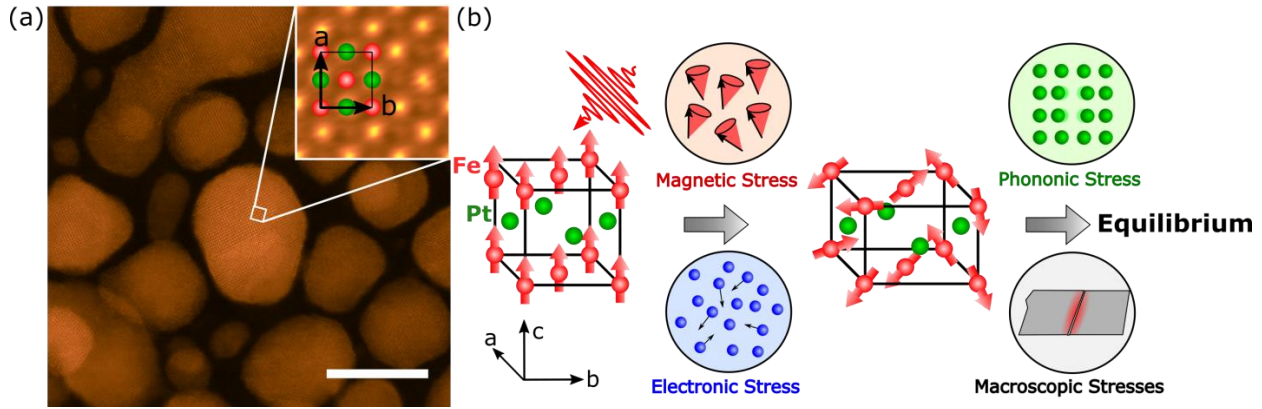

**Fig. 1** FePt physical structure and the time sequences of contributions to the lattice strain. **a** High-angle annular dark field scanning transmission electron microscopy images of the morphology and *a*-, *b*-plane atomic arrangement (inset) for FePt nanoparticles embedded in an amorphous carbon matrix. 10 nm scale bar shown in lower right. **b** The FePt crystal structure and magnetic arrangement. Fe and Pt atomic positions in the $L1_0$ lattice structure are indicated by red and green spheres in the insets, respectively. The majority of the grains are aligned with the *a* and *b* unit cells directions in the plane of the film. The schematic shows the sequence of stress contributions in FePt. In the first picosecond after laser excitation the lattice is subject to the rapidly developing magnetic and electric stresses. This leads to an increase in the tetragonal distortion of the unit cell as illustrated. On a timescale of a few picoseconds, phonon and macroscopic stresses become important and the lattice moves towards its equilibrium state.

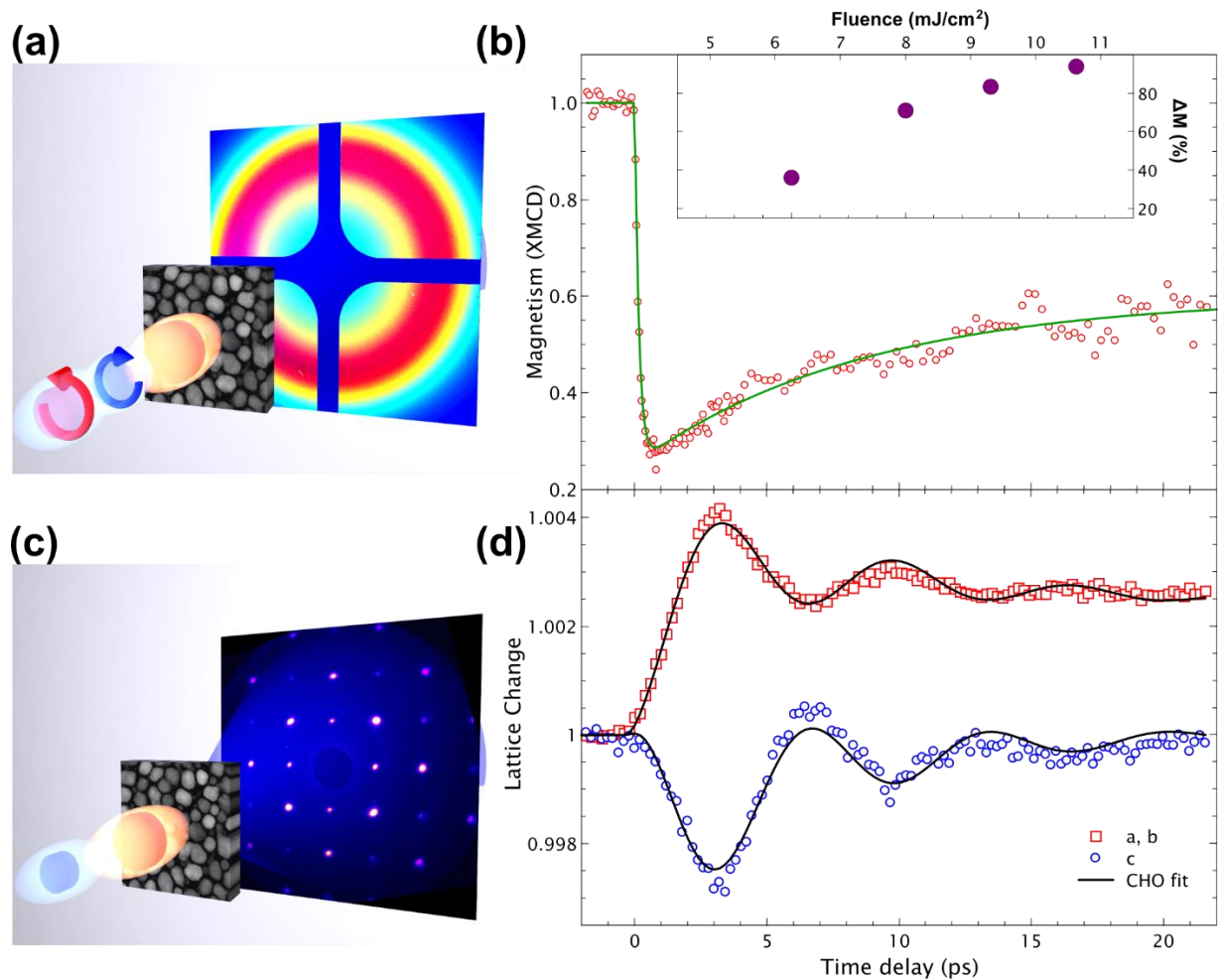

**Fig. 2** Ultrafast spin and lattice dynamics of the FePt nanoparticles. **a** Experimental setup for time-resolved soft X-ray scattering measurements of the FePt magnetization dynamics. **b** The measured change in magnetization for a laser pump fluence of 9.5 mJ/cm$^2$. The solid line is a two-exponential fit to the data showing a 146 ± 15 fs drop followed by a 16 ± 4 ps recovery of the magnetization. The inset shows the size of the magnetization drop *vs.* pump fluence. **c** Experimental setup for ultrafast electron diffraction measurements of the time-resolved structural response of the FePt nanoparticles. **d** Measurements of the Bragg diffraction are used to determine the temporal evolution of the *a*-, *b*-lattice (red squares) and *c*-lattice (blue circles) spacings for a pump laser fluence of 5.0 mJcm$^{-2}$. The solid lines are fits of a simple driven coupled harmonic oscillator (CHO) model to the data as described in the text.

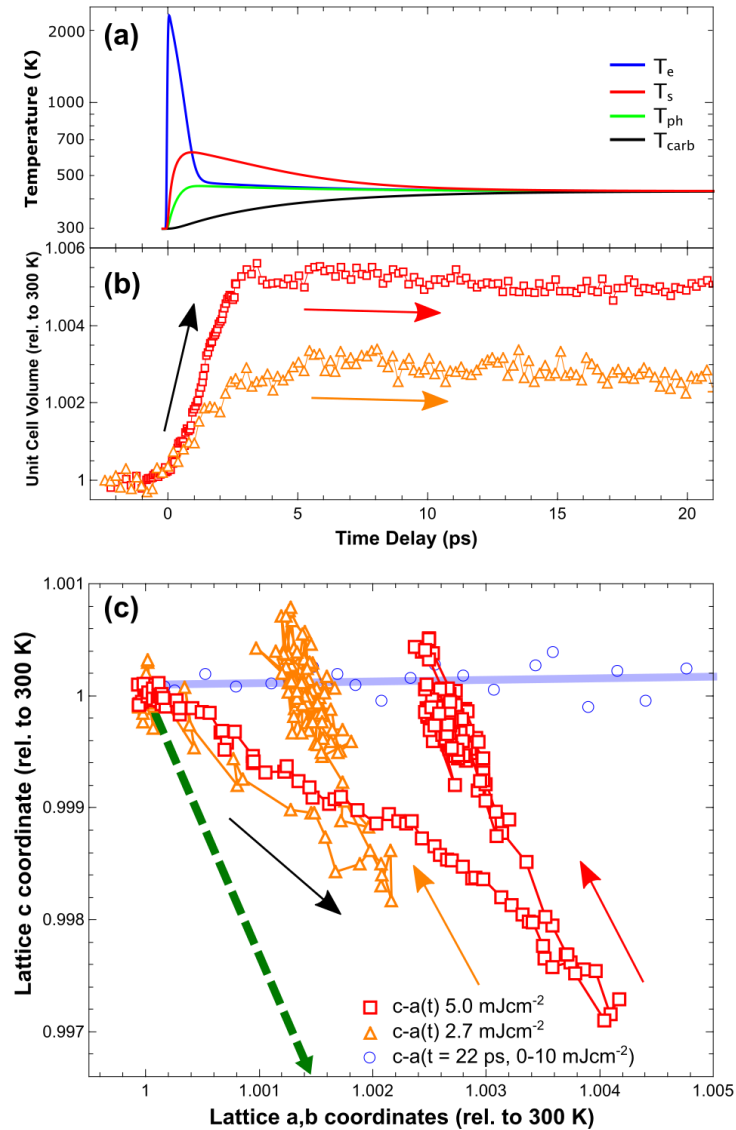

**Fig. 3** FePt, sublattice temperature, volume change and evolution of the FePt $a$–$c$ lattice parameters. **a** An extended three-temperature model of electronic, spin and lattice temperature evolution in FePt, together with the lattice temperature of the carbon matrix upon pumping with a laser fluence of 5.0 mJcm$^{-2}$. **b** The temporal evolution of the unit cell volume ($V = c \cdot a^2$) calculated from the $a$ & $c$ lattice data at laser fluences of 5.0 mJcm$^{-2}$ (red) and 2.7 mJcm$^{-2}$ (orange). **c** The FePt lattice motion plotted as a relative change in the $c$–$a$ plane, with arrows indicating the time direction. Quasi-equilibrium measurements of the nanoparticle thermal expansion as a function of laser fluence, made at a time delay of 22 ps, are shown as blue open circles. The dashed green arrow indicates the direction of the $a$- and $c$-lattice evolution corresponding to the calculated change between FM to PM ground states.

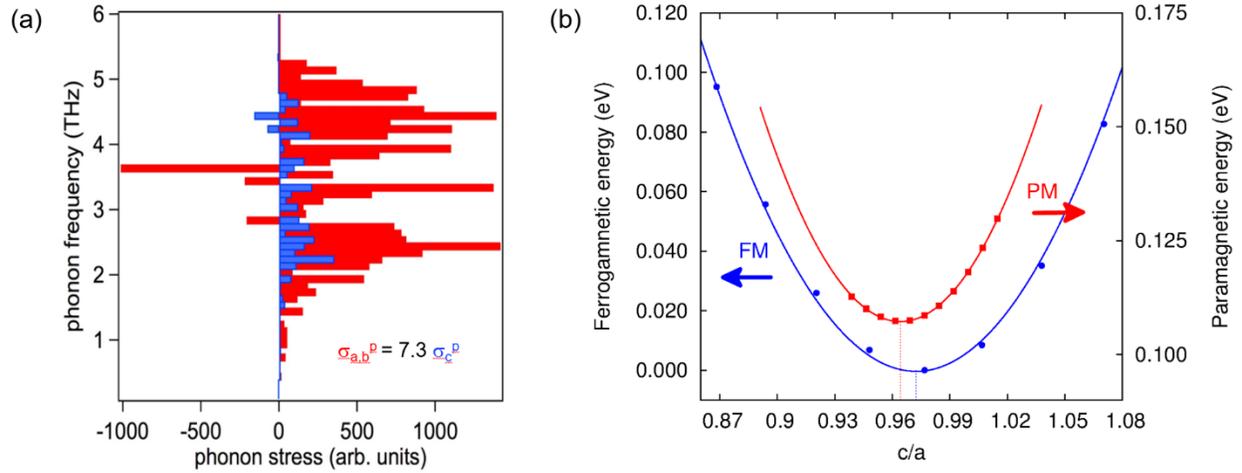

**Fig. 4** *Ab-initio* calculations of the lattice and magnetic stress. **a** The mode resolved lattice stress, $\sigma_i^{ph}$, along lattice directions, $i = a$ and $c$, calculated *ab initio* for a non-equilibrium phonon population. **b** The *ab initio* calculated total energies versus $c/a$ ratio for FePt in the ferromagnetic (FM, blue circles) and paramagnetic (PM, red circles) phase at constant volume. Calculations in the full $a$-, and $c$-plane (not shown) determine the lattice parameters change to $\Delta a/a = +0.25$ % and $\Delta c/c = -0.53$ % when moving from the global energy minimum (not shown) of the ferromagnetic to the paramagnetic phase. We note that the calculated ferromagnetic ground-state value of $c/a = 0.964$ is in good agreement with $c/a = 0.972 \pm 0.003$ measured at room temperature (see Supplementary Figure 3c).

# Tables

Table 1: Parameters determined from fitting the coupled harmonic oscillator to the data in Fig 2d.

| Stress | Equation/Model | Fitting value | ab coefficient | c coefficient |
|---|---|---|---|---|
| $\sigma^e$ | $\sigma^e T_e(t)\delta T_e(t)$ | 8.3e-10 ps$^{-2}$K$^{-2}$ | 1 | 0.5 |
| $\sigma^{ph}$ | $\sigma^{ph}[1 - \exp(-\frac{t}{\tau_l})]$ | 3.9e-3 ps$^{-2}$ | 1 | 0.14 |
| $\sigma^m$ | $\sigma^m \left(\frac{\Delta M(t)}{M_0}\right)^2$ | 9.6e-3 ps$^{-2}$ | 0.47 | -1 |
| $\sigma^{carb}$ | $\sigma^{carb} T_{carb}(t)$ | 1.5e-5 ps$^{-2}$K$^{-1}$ | 1 | 0 |

Table 2: Model parameters used for the four-temperature model.

| Bath\Parameter | $C_P$ (J mol$^{-1}$K$^{-1}$) | Coupling (10$^{18}$ Wm$^{-3}$K$^{-1}$) | Excitation Energy (J) |
|---|---|---|---|
| FePt electronic | $C_e = (4.0e-3)T_e$ | $g_{es} = 0.08$ <br> $g_{eph} = 0.7$ | $5e-4$ <br> ($5\ mJcm^{-2}$) |
| FePt spin | $C_s(300K) = 1.5$ <br> $C_s(600K) = 4.1$ <br> $C_s(700K) = 10.6$ <br> $C_s(750K) = 34.7$ <br> $C_s(755K) = 0.3$ | $g_{es}$=0.08 | - |
| FePt phonon | $C_{ph} = 71.4$ | $g_{eph} = 0.07$ <br> $g_{phc} = 1.6$ | - |
| Carbon Matrix | $C_{carb} = 8.5$ | $g_{phc} = 1.6$ | - |